\documentclass[final,english]{bullsrsl}[2022/06/15]

\usepackage[latin1]{inputenc}
\usepackage[T1]{fontenc}

\usepackage{natbib} 
\usepackage{graphicx}
\usepackage[flushleft]{threeparttable}

\hypersetup{
        colorlinks   = true, 
        urlcolor     = magenta, 
        linkcolor    = blue, 
        citecolor   = blue 
}

\newcommand{\hrr}{HR~8038}
\newcommand{\ach}{AC~Her}
\newcommand{\hdd}{HD~76446}

\newcommand\arcs{\hbox{$^{\prime\prime}$}}    

\begin{document}
\title{Near-UV and optical spectroscopic investigation of late-type\\ stars from MIRA/\emph{Oliver Observing Station}}

\author[affil={1}, corresponding]{Subhajeet}{Karmakar}
\author[affil={2}]{Avrajit}{Bandyopadhyay}
\author[affil={1}]{Wm. Bruce}{Weaver}
\author[affil={1,3}]{Riddhi}{Shedge}
\author[affil={4}]{Jeewan C.}{Pandey}
\author[affil={1,5}]{Daniel V.}{Cotton}
\author[affil={1}]{Jean}{Perkins}

\affiliation[1]{Monterey Institute for Research in Astronomy (MIRA), 200 Eighth Street, Marina, California 93933, USA}
\affiliation[2]{University of Florida, Gainesville, Florida, 32611, USA}
\affiliation[3]{Monta Vista High School, 21840 McClellan Rd, Cupertino, California 95014, USA}
\affiliation[4]{Aryabhatta Research Institute of Observational Sciences (ARIES), Manora Peak, Nainital 263002, India}
\affiliation[5]{Western Sydney University, Locked Bag 1797, Penrith-South DC, NSW 1797, Australia}
\correspondance{subhajeet09@gmail.com, sk@mira.org}
\date{13th October 2020}
\maketitle

\begin{abstract}
Late-type stars are the most abundant in the galactic stellar population. These stars, with the similar internal structure to the Sun, are expected to have solar-like atmospheres.
Investigating the stellar parameters and chemical abundances on late-type stars is essential to provide valuable constraints about stellar age, chemical evolution, and atmosphere of exoplanets. 
In this work, we present the study of the Near-UV and optical spectroscopic observation of three late-type stars: \hrr, \ach, \hdd, as obtained from 36-inch MIRA/\textit{Oliver Observing Station}.  
We derived surface temperature, gravity, metallicity, and the chemical abundances of light element Carbon in the stellar atmosphere. The elemental abundance of the Carbon for HR~8038, AC~Her, and HD~76446 are derived to be 95\%, 97\%, and 108\%, respectively, of the solar value.  
\end{abstract}


\section{Introduction}
\label{sec:intro}
Most of the stellar population of our Galaxy consists of late-type stars \citep{Bochanski-10-AJ-2}. With partially- or fully-convective envelopes, late-type stars are expected to have solar-like atmospheres and magnetic activities \citep[][]{Priest-00-eaa-8, KarmakarS-19-PhDT-2}. However, the stars present a wide range of temperature, gravity, metallicity, and elemental abundances that have been found to vary significantly over time \citep[][]{JofreP-14-A+A-3, BandyopadhyayA-22-ApJ, KarmakarS-16-MNRAS-2, KarmakarS-17-ApJ, KarmakarS-19-BSRSL-2, KarmakarS-20-MNRAS-3, KarmakarS-22-MNRAS-1, KarmakarS-23-MNRAS-2}. Therefore, a robust observational and theoretical investigation is essential. 
Investigating the elemental abundances of late-type stars provides key information for characterizing the stellar population of our Galaxy. Carbon being the fourth most abundant element in the universe (after hydrogen, helium, and oxygen), is of particular interest in many fields of astrophysics, including stellar age determination \citep[][]{BondH-13-ApJ-2, RomanoD-20-A+A-4, ZhangX-21-RAA-50, BeverageA-23-ApJ}, chemical evolution of galaxies \citep[][]{ChiappiniC-03-A+A-4, CarigiL-05-ApJ, CescuttiG-09-A+A, BotelhoR-20-MNRAS, GustafssonB-22-Univ}, and structure of exoplanets \citep[][]{BondJ-10-ApJ-4, MadhusudhanN-12-ApJ-6, PelletierS-21-AJ-2, GrantD-23-ApJ-3}.

For the past two decades, the Monterey Institute for Research in Astronomy (MIRA) has observed hundreds of spectra of low-mass stars. In this paper, utilizing a subsample of this survey, we present an investigation of surface temperature, gravity, metallicity, and Carbon abundances using the NUV and optical low-resolution spectra of three late-type stars: HR~8038, AC~Her, and HD~76446. These F and G-type stars are located at a distance of 56.4$\pm$1.6, 1231$\pm$44, and 81.5$\pm$0.9~pc, respectively \citep[][]{Bailer-Jones-18-AJ-6}. Some basic information of these stars are shown in Table~\ref{tab:obslog}. We structured the paper as follows: Section~\ref{sec:redn} describes the observations and data analysis. Section~\ref{sec:modeling} describes the models and methods utilized for atmospheric modeling and presents the results. Finally, in Section~\ref{sec:concl}, we discuss the result and outlined the conclusions drawn from our study.

\section{Observations and Data Analysis}
\label{sec:redn}
The low-resolution ($\sim$5,000) spectroscopic observations of three late-type stars have been carried out using the 36-inch F/10 classical Cassegrain Telescope mounted at the \textit{Bernard M. Oliver Observing Station} \citep[OOS;][]{WeaverW-75-BAAS-2} of the \textit{Monterey Institute for Research in Astronomy} (MIRA). The OOS is located at an elevation of 5010 feet on Chews Ridge in California -- one of the best-seeing sites in the Pacific region \citep[$\approx$1\arcs\!\!.1;][]{HutterD-97-AJ-3, WalkerM-70-PASP-2}. 
For all three observations, the dual-port spectrograph \citep[DPS; see][]{Torres-dodgenA-93-PASP} has been used with a 1200 lines per mm grating along with a back-illuminated 1024~$\times$~255  ANDOR Camera \citep[Model: DU420-BU; see][]{WalkerR-07-Icar-1}. The observations were performed within the wavelength regions 3500--6800~\AA, with exposure times between 420 and 1800~s. Detailed information on the objects and specific observations is given in Table~\ref{tab:obslog}.

\begin{table}[h]
\vspace{-0.5cm}
	\centering
    \begin{minipage}{150mm}
	\caption{Observation details of the Program stars.} 
	\label{tab:obslog}
    \end{minipage}
    \tabcolsep=0.24cm
	\begin{tabular}{lccccccc}
      \hline\hline\\[-5mm]
    Object   &  SpT &  RA            &	Dec           & Obs-Date      & Exp Time (s) & V (mag)\\[-0mm]
    \hline \hline \\[-5mm]
    HR~8038  &  F1 & 21 00 03.99 &  +07 30 58.28  &  2017-10-23    & 720  & 5.99  \\
    AC~Her   &  F4 & 18 30 16.24 &  +21 52 00.60  &  2010-06-15    & 420  & 7.01  \\
    HD~76446 &  G2 & 08 56 33.11 &  +12 25 54.78  &  2010-02-17    & 1800 & 8.42  \\
	\hline
    \normalsize
	\end{tabular}
\vspace{-0.5cm}
\end{table}

\begin{figure*}[t]
\centering
\includegraphics[width=16cm, angle=-0, trim={1mm 11.7mm 1.5mm 0mm},clip]{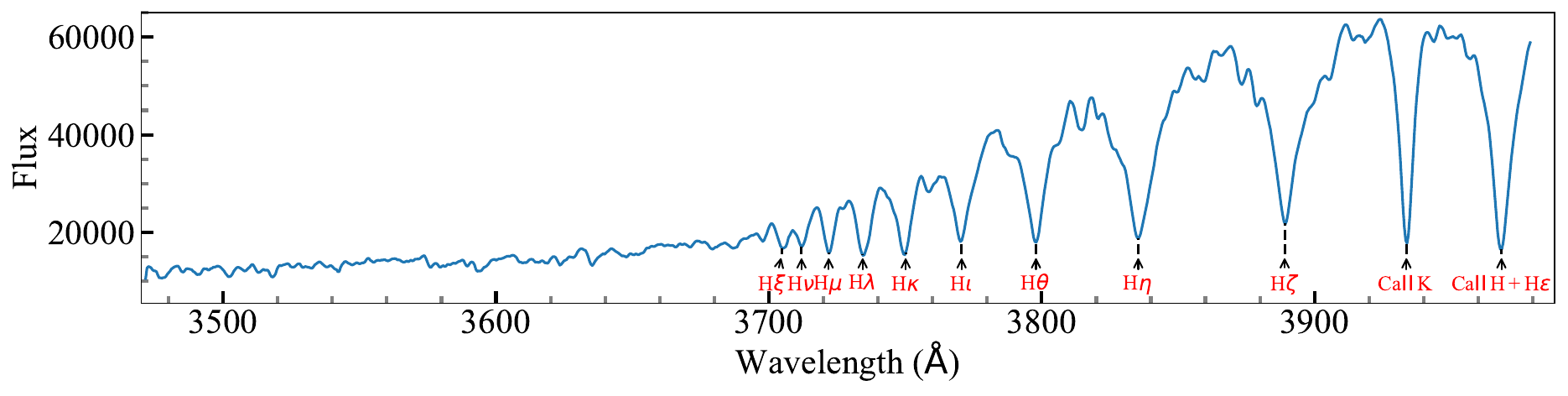}
\includegraphics[width=16cm, angle=-0, trim={1mm 11.7mm 1.5mm 0mm},clip]{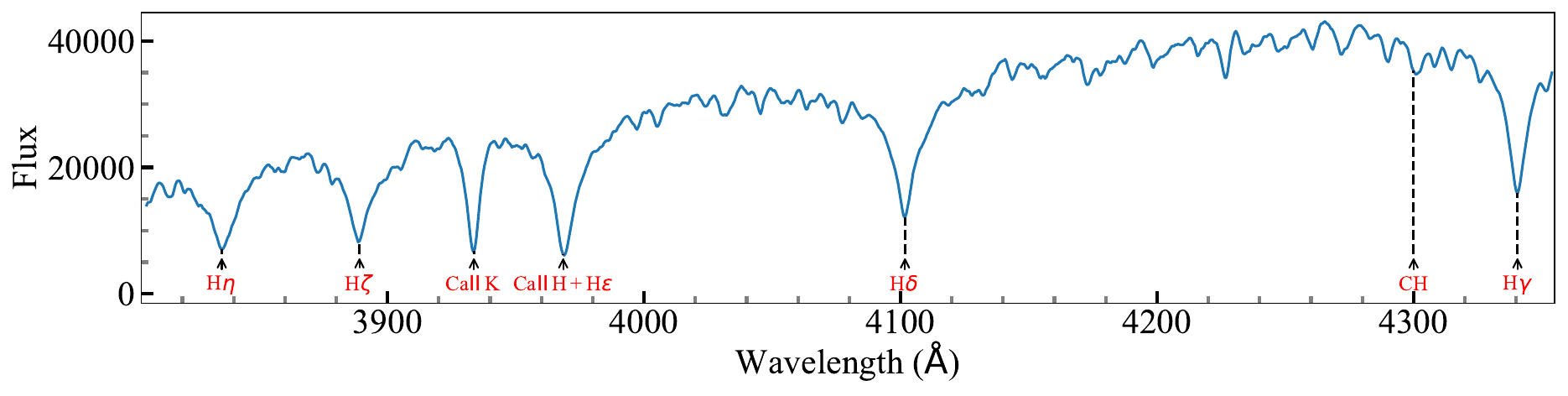}
\includegraphics[width=16cm, angle=-0, trim={1mm 11.7mm 1.5mm 0mm},clip]{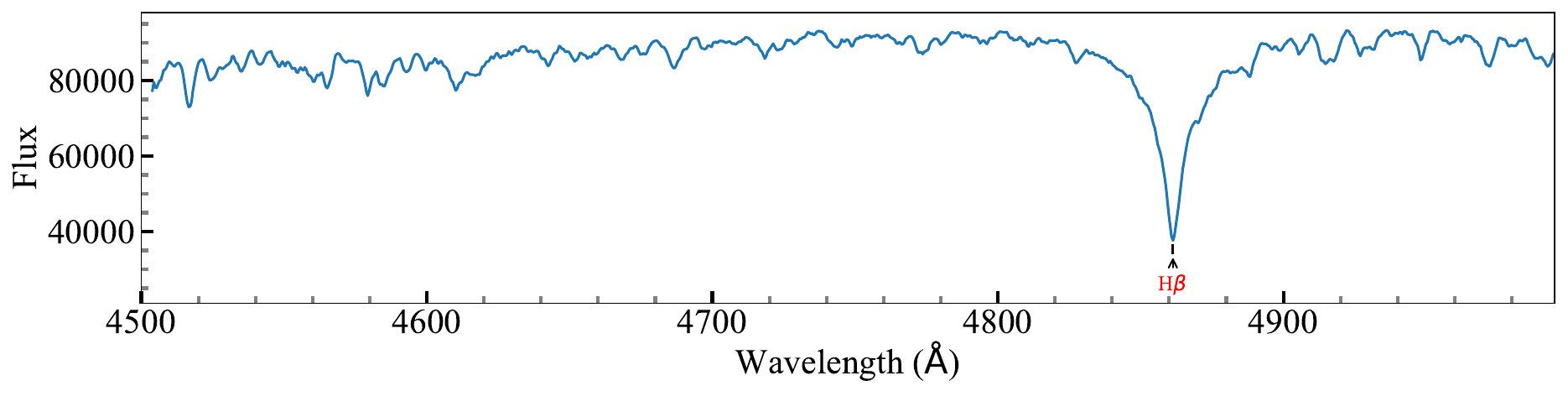}
\includegraphics[width=16cm, angle=-0, trim={1mm 4mm 1.5mm 0mm},clip]{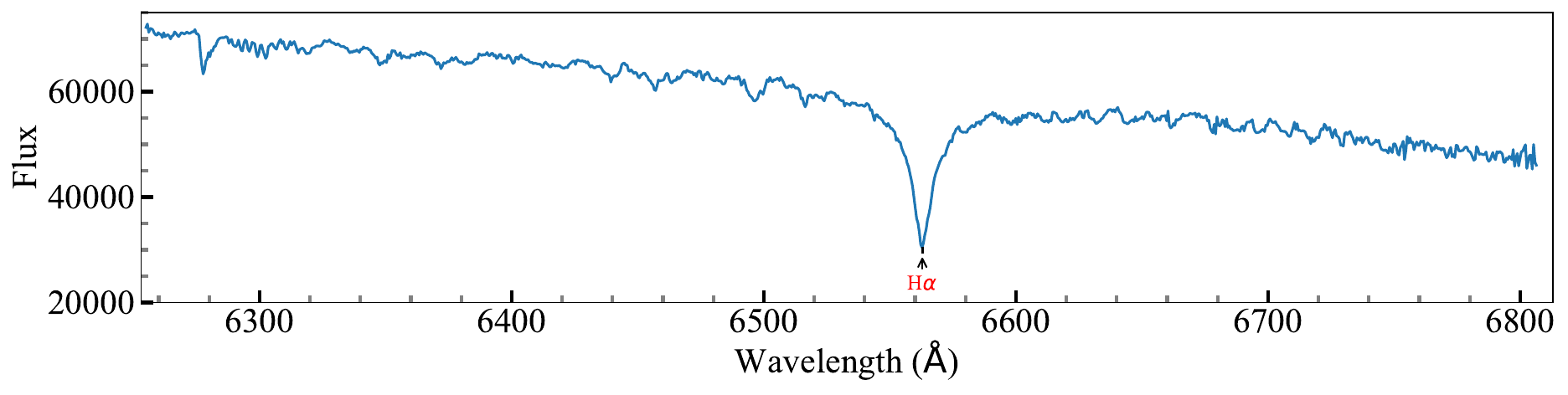}
\bigskip
\begin{minipage}{12cm}
\caption{The wavelength-calibrated NUV and optical spectra of HR 8038, along with identified strong hydrogen and calcium lines, are shown as representative spectra.}
\label{fig:input_hr8}
\end{minipage}
\end{figure*}

In this research, the data reduction was performed using the \textit{Image Reduction and Analysis Facility} \citep[\href{https://iraf.net/irafdocs/}{\tt iraf};][]{TodyD-86-SPIE} software. In order to perform various operations, including initial bias, cosmic ray, and flat field corrections, identification of spectral lines, and wavelength calibrations, we used {\tt iraf} packages, \textit{viz.}, {\tt imutil}, {\tt crutil}, {\tt noao.imred}, {\tt noao.onedspec}, and {\tt noao.twodspec}. For further operations with wavelength-calibrated spectra, we have also used {\tt Python} packages, including \href{https://matplotlib.org/stable/index.html}{\tt matplotlib} \citep[][]{HunterJ-07-CSE-1}, and \href{https://github.com/astropy/astropy}{\tt astropy} \citep[][]{Astropy-Collaboration.-22-ApJ}. A representative wavelength calibrated spectra of \hrr, as observed on October 23, 2017, is shown in Figure~\ref{fig:input_hr8}.

\section{Spectral Fitting \& Atmospheric Modeling}
\label{sec:modeling}
In order to investigate the photospheric absorption features, we initially computed steady-state 1-D plane-parallel model atmospheres in local thermodynamic equilibrium (LTE) with the \href{https://wwwuser.oats.inaf.it/castelli/sources/atlas9codes.html}{\tt ATLAS9} model \citep[][]{KuruczR-93-PhST-21, CastelliF-03-IAUS-1}.
To determine the stellar metallicity and detailed abundances, we used a realistic non-LTE version of the widely used state-of-the-art \href{https://ascl.net/1205.004}{\tt TURBOSPECTRUM} synthesis code \citep[][]{AlvarezR-98-A+A-2, PlezB-12-ascl, GerberJ-23-A+A}.

\begin{figure*}
\centering

\includegraphics[width=7.9cm, angle=-0]{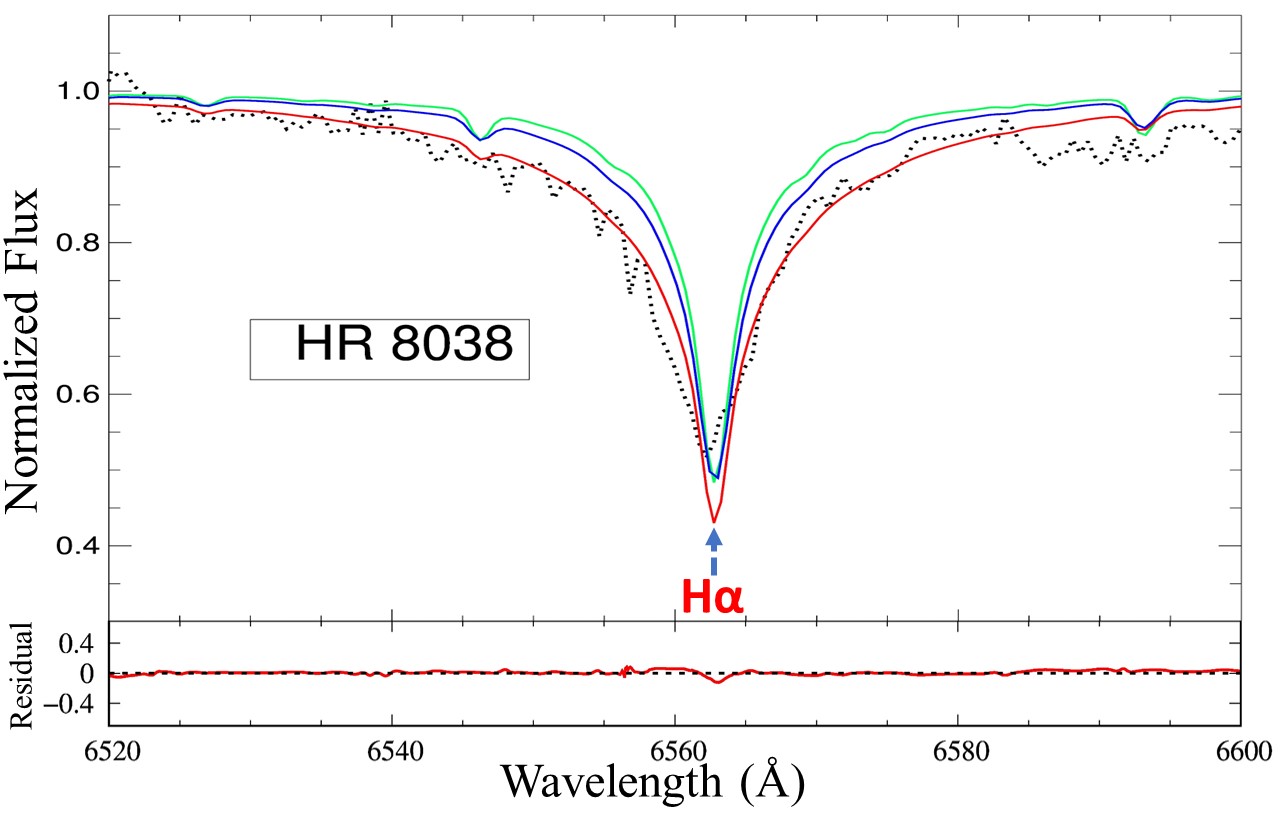}
\includegraphics[width=7.9cm, angle=-0]{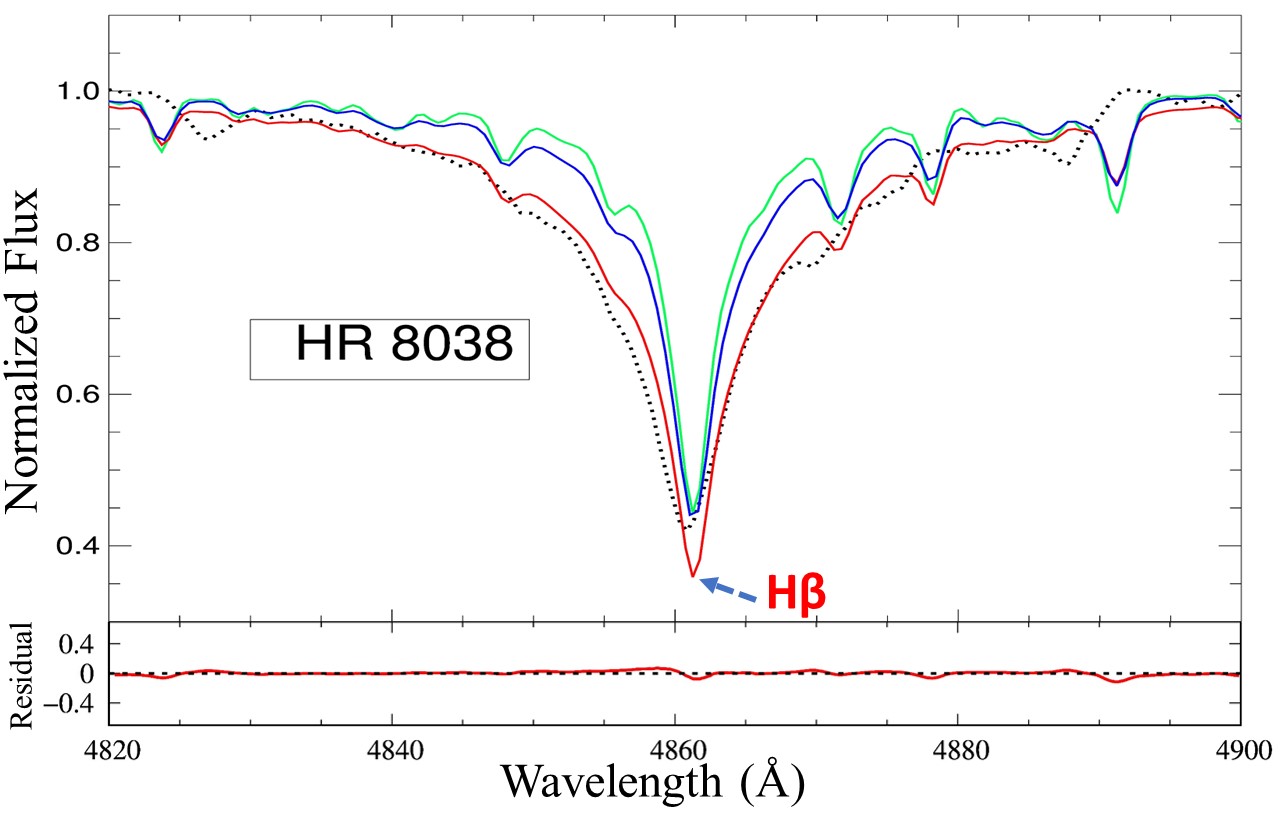}
\includegraphics[width=7.9cm, angle=-0]{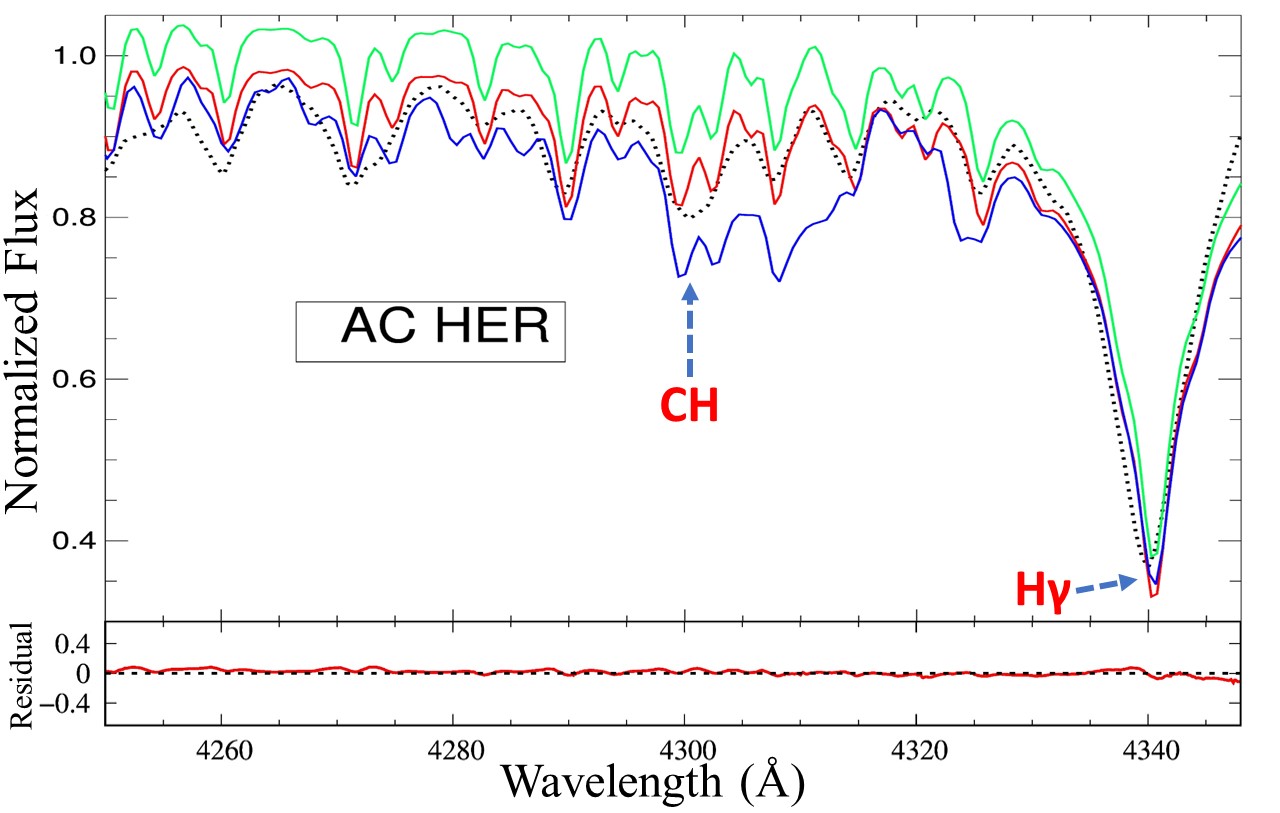} 
\includegraphics[width=7.9cm, angle=-0]{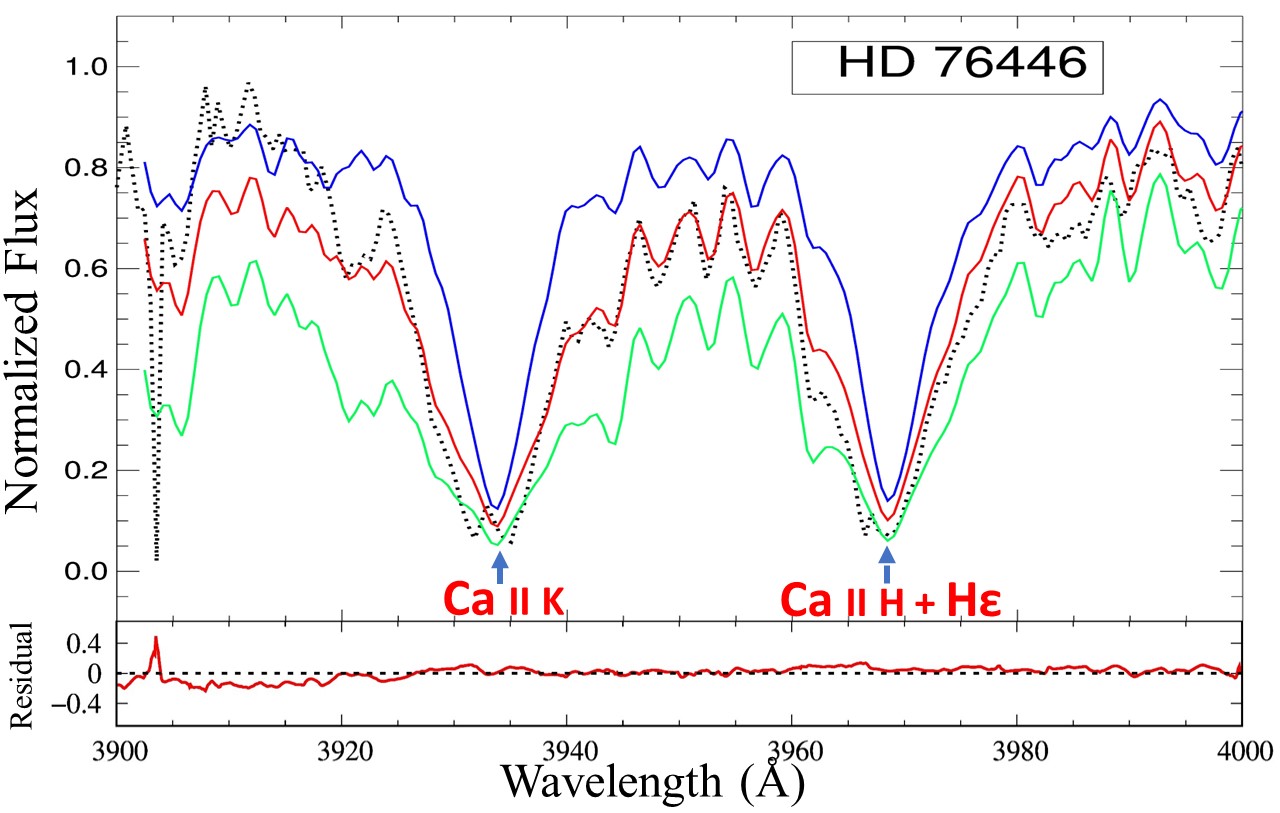}
\bigskip
\begin{minipage}{12cm}
\caption{The spectral fitting of a few representative spectra is shown. The upper panel of each figure shows the observed spectra (black dotted line), best-fitted spectra (red solid line), and the typical variations from the best fit for a temperature and log~g of 300~K and 1.0~dex (green and blue solid lines). The lower panel of each figure shows the residual of the observed and best-fitted spectra. In the top panel of each figure, the identified absorption lines have been marked with red-colored text and blue arrows, whereas the black label indicates the name of the corresponding low-mass star.}
\label{fig:modeled}
\end{minipage}
\end{figure*}

We estimate the stellar parameters using the broad and strong H-features and Ca~{\sc ii}~H and K lines by iteratively varying the stellar parameters to arrive at the best fits. Figure~\ref{fig:modeled} shows a few representative absorption features of the program stars. The black dotted lines are the observed spectra, and the solid red lines show the best fit. The typical variations from the best-fit for a temperature variation of 300~K and log~g variations of 1.0 dex are shown using green and blue solid lines, whereas the residuals of the observed and the best-fitted spectra are shown in the bottom panels.
In the representative spectra shown in Figure~\ref{fig:modeled}, we have marked the identified absorption features of H$\alpha$,  H$\beta$, Ca~II~H (along with H$\epsilon$), Ca~II~K, H$\gamma$, and CH molecular band. The estimated best-fitted parameters of stellar surface temperature (T$_{eff}$), surface gravity (log~$g$), and metallicity ([Fe/H]) are given shown in Table~\ref{tab:params}. We also estimated the Carbon abundance A(C) from the absorption line at 4300 \AA. The best-fitted Carbon abundance is shown in the rightmost panel of Table~\ref{tab:params}.

\section{Discussion \& Conclusions}
\label{sec:concl}
In this paper, we present an analysis of stellar parameters using low-resolution Near-UV and optical spectra. We estimated the surface temperatures of HR~8038, AC~Her, and HD~76446 to be 7100, 5650, and 5850~K, respectively. In the literature, there are 17, 12, and 8 estimations of surface temperatures available for these three sources.
The known surface temperatures of \hrr, \ach, and \hdd\ are found to be within the ranges of 7051 -- 7850, 5080 -- 5933, and 5730 -- 6027~K, respectively. The surface gravities, on the other hand, are previously derived to be within the ranges of 3.69 -- 3.99, 1.39 -- 3.79, and  3.85 -- 4.13, whereas the metallicities were estimated to be in the ranges of -1.08 to -0.46, -1.35 to -0.75, and -0.58 to +0.03, respectively. We found that seven out of nine parameters estimated from the spectral fitting are within the 1$\sigma$ uncertainty level. Although the objects considered for this paper are well-studied, obtaining similar values for the stellar parameters using low-resolution spectra is an interesting finding. However, the only source \hrr\ shows a slight deviation in T$_{eff}$ and a significant deviation in log~$g$, and it needs further investigation.

In order to estimate the Carbon abundances, we used the best-fit values for T$_{eff}$, log~$g$, and [Fe/H]. The Carbon abundances A(C) of HR~8038, AC~Her, and HD~76446 are estimated to be 8.0, 8.15, and 9.1. Considering the solar Carbon abundance value of 8.39 \citep{GrevesseN-07-SSRv-2}, we found the Carbon abundance of HR~8038, AC~Her, and HD~76446 are 95\%, 97\%, and 108\% of the solar value. We expect to report more results from the low-mass spectroscopic survey of MIRA with a robust investigation of spectral lines in the future. The analysis with the complete sample of 130 low-mass stars will provide useful information on stellar abundances and will further enable us to understand the First Ionization Potential effect on low-mass stars. This study will provide important information to understand the stellar magnetic dynamo and chemical evolution of the Galaxy.

\begin{table}[t]
\vspace{-0.3cm}
 \begin{threeparttable}
	\centering
    \begin{minipage}{150mm}
	\caption{Estimated stellar parameters from the spectra.} 
	\label{tab:params}
    \tabcolsep=0.64cm
	\begin{tabular}{lcccc}
      \hline\hline\\[-5mm]
      Object & T$_{eff}$ &	log~$g$  & [Fe/H] & A(C)$_{\star}$$\ddagger$ \\[-0mm]
    \hline \hline \\[-5mm]
    HR~8038  &   7100$\pm$300~K  &   4.50$\pm$0.5    & --0.8$\pm$0.3  & --0.39$\pm$0.50  \\
    AC~Her   &   5650$\pm$300~K  &   2.10$\pm$0.5    & --1.3$\pm$0.3  & --0.24$\pm$0.45 \\
    HD~76446 &   5850$\pm$300~K  &   4.00$\pm$0.5    & --0.5$\pm$0.3  & 0.71$\pm$0.40  \\
	\hline
    \normalsize
	\end{tabular}
\vspace{-0.7cm}
   \begin{tablenotes}
      \small
    \item$\ddagger$ -- The solar Carbon abundance A(C)$_{\odot}$ is adopted from \citep{GrevesseN-07-SSRv-2}, whereas the stellar abundance A(C)$_{\star}$ is estimated as A(C)$_{\star}$ = A(C) -- A(C)$_{\odot}$. 
    \end{tablenotes}
    \end{minipage}
  \end{threeparttable} 
\end{table}

\vspace{-3mm}
\begin{acknowledgments}
  This research is based on the observations obtained with the Dual-Port Spectrograph instrument at MIRA/OOS. We acknowledge \textit{Friends of MIRA} for supporting this research. This research uses {\tt ATLAS9} atmospheric model and {\tt TURBOSPECTRUM} spectral synthesis code. We also acknowledge the high-performence research computing facility at University of Florida.
\end{acknowledgments}
\vspace{-3mm}
\vspace{-3mm}
\begin{furtherinformation}
\begin{orcids}
\orcid{0000-0001-8620-4511}{Subhajeet}{Karmakar}
\orcid{0000-0002-8304-5444}{Avrajit}{Bandyopadhyay}
\orcid{0009-0005-0584-3343}{Wm. Bruce}{Weaver}
\orcid{0000-0002-2489-5908}{Riddhi}{Shedge}
\orcid{0000-0002-4331-1867}{Jeewan C.}{Pandey}
\orcid{0000-0003-0340-7773}{Daniel V.}{Cotton}
\orcid{0000-0002-6703-5406}{Jean}{Perkins}
\end{orcids}
\vspace{-3mm}

\begin{authorcontributions}
  \textbf{SK:} Conceptualization, Project administration, Formal analysis (lead), Investigation (lead), Supervision (mentored intern RS), Software ({\tt IRAF}, {\tt Python}, {\tt SHELL}), Visualization, Writing - Original Draft; 
  \textbf{AB:}  Software ({\tt IDL}, {\tt ATLAS9}, {\tt TURBOSPECTRUM}), Formal analysis (spectral fitting), Investigation (atmospheric modeling), Visualization, Writing - Review \& Editing;
  \textbf{WBW:} Investigation (observation), Writing - Review \& Editing;
  \textbf{RS:} Software ({\tt IRAF}, {\tt SHELL}), Formal analysis (data reduction), Investigation (wavelength calibration);
  \textbf{JCP:} Investigation (stellar absorption features), Writing - Review \& Editing;
  \textbf{DVC:} Writing - Review \& Editing.
  \textbf{JP:} Writing - Review \& Editing;
\end{authorcontributions}
\vspace{-3mm}

\begin{conflictsofinterest}
The authors declare no conflict of interest.
\end{conflictsofinterest}

\end{furtherinformation}

\bibliographystyle{bullsrsl-en}

\bibliography{SK_collections}

\end{document}